\documentstyle[11pt,paspconf,epsf,twoside]{article}

\begin{document}

\title{Galaxies at $z>5$: The View from Hawaii} 

\author{Esther M. Hu and Lennox L. Cowie}
\affil{Institute for Astronomy, University of Hawaii, 2680 Woodlawn Drive,
    Honolulu, HI 96822, USA}
\author{Richard G. McMahon}
\affil{Institute of Astronomy, Madingley Road, Cambridge CB3 0HA, UK}

\begin{abstract}
We review the properties of $z>5$ galaxies studied with {\it HST\/}
and with the Keck telescopes, and discuss the detectability of Ly$\alpha$ 
emission-line galaxies out to $z\sim6.5$ based on these data and ongoing 
narrowband imaging surveys.  The brightest sources may show $(R-Z)$ color 
breaks, although the high sky background at $Z$ ($\lambda_{\rm eff}\approx 
9200$ \AA), makes such observations challenging for typical faint sources.
Keck LRIS observations of the $z=5$ SDSS quasar and $z>5$ galaxies observed with
{\it HST\/} in the HDF show that the strength of the Lyman break is 
evolving more slowly than extrapolations from models at $z\sim3$ would
predict.
\end{abstract}


\section{Introduction}

Over the last year and a half, a number of very high redshift ($z>5$)
galaxies have been reported (Dey et al.\ 1998, Hu et al.\ 1998, Weymann et
al.\ 1998, Spinrad et al.\ 1998, Chen et al.\ 1999, Hu et al.\ 1999), and our
knowledge of the $z>4$ galaxy population (e.g., Hu et al.\ 1996, Petitjean et
al.\ 1996, Fontana et al.\ 1996, Hu \& McMahon 1996, Trager et al.\ 1997,
Franx et al.\ 1997, Hu et al.\ 1997, Frye \& Broadhurst 1998, Hu et
al.\ 1998, Soifer et al.\ 1998, Pell\'o et al.\ 1999, Steidel et al.\ 1999,
Hu et al.\ 1999) has been substantially increased. These samples have
recently been joined by radio galaxy identifications which reach
$z>5$ (van Breugel et al.\ 1999a, 1999b). At the highest redshifts
now probed, identification of galaxies basically relies on two key
diagnostics:  the Lyman break across the continuum and the redshifted
Ly$\alpha$ line.  Both characteristics have been used to select and to verify
high-$z$ galaxy candidates.  

The extremely faint nature of the high-$z$ galaxy population 
in infrared continuum light and the increasing strength and frequency
of nightsky emission lines at very long wavelengths makes it clear that
the contrast and detection of high-redshift galaxies from the ground
is challenging, even for the new generation of 8- to 10-m class telescopes.  
The discovery of very high equivalent width emitters at $z=4.55$ 
(Hu \& McMahon 1996)
indicated the existance of a substantial population of strong Ly$\alpha$
emitters which could be observed to very high redshifts, beyond $z>5$. 
In this paper we report on the status of ongoing Ly$\alpha$ emission-line
searches combined with color-selection data and consider the prospects for
future identifications at the very high redshift end.

\section{Ly$\alpha$ Searches}

\begin{figure}
\plotone{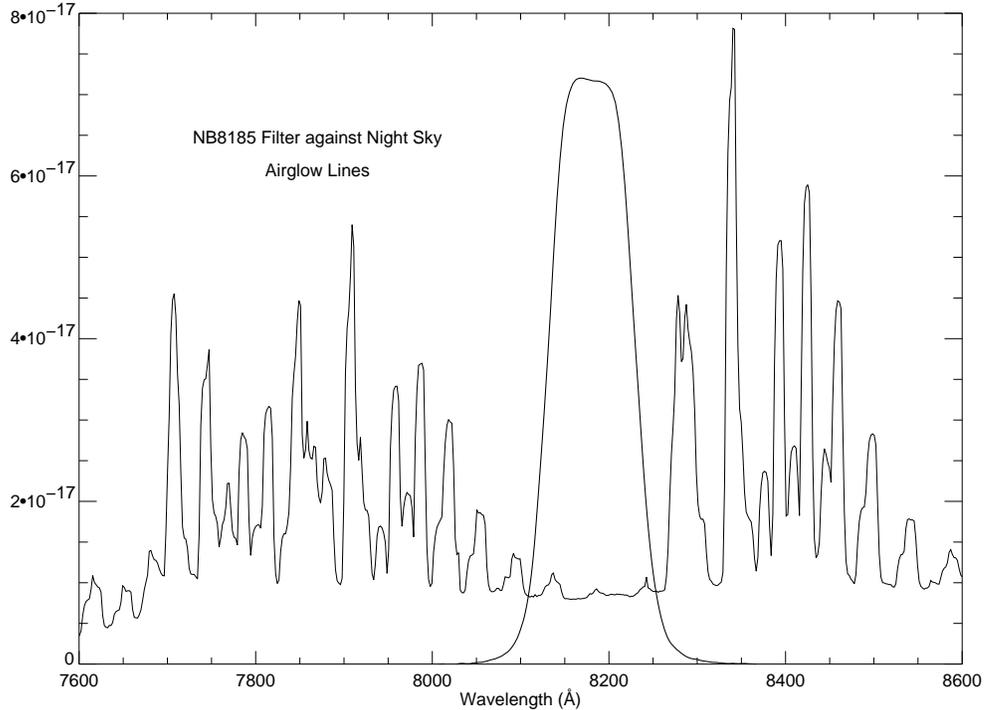}
\caption{\small Narrowband filter at 8185 \AA\ and fluxed background night
sky lines.}
\label{fig:0}
\end{figure}

The Ly$\alpha$ searches are designed to address issues of detectability of
field galaxies at very high redshifts, using the increased contrast of the
object against the background sky when viewed in the Ly$\alpha$ emission
line.  The contrast of emission features against the neighboring continuum is
also enhanced by the $(1+z)$ magnification of line widths, producing high
equivalent width signatures.  Through the use of spectral regions free of
strong night sky lines (e.g., Fig.~\ref{fig:0}), these studies are also designed to address object
statistics over a range of different redshift intervals, since our ability to
detect and reliably confirm very high-redshift galaxies is otherwise highly
dependent on the location of emission features with respect to
background nightsky lines.  Narrowband Ly$\alpha$ searches combined with deep
multi-color data on the Hawaii Survey Fields and HDF have been used as a
training set (Cowie \& Hu 1998; Hu 1998) to test emission line
identifications in terms of color and equivalent width diagnostics over an
extensive spectroscopic database, starting at $z\sim3.4$, where color
selection is robust, and working out to successively higher redshifts ($z\sim
4.6, 5.7, 6.5$).  This allows studying continuity in properties of genuine
Ly$\alpha$ emitters and both color-selected galaxies and low-redshift
foreground emitters at progressively longer wavelengths, where higher sky
backgrounds from the increased density of strong nightsky lines and fainter
continua for the more distant galaxies are issues.  A variant approach is to
use a Fabry-Perot (Calar Alto Deep Imaging Survey, Meisenheimer et al.\
1998), which requires a high-level of precision in low-level light processing
and flat-fielding. The present Ly$\alpha$ surveys have reached $z\sim 5.7$
(Hu et al.\ 1999), and $z\sim6.5$ narrowband studies have recently been
started.

From the initial wide-field narrowband surveys for Ly$\alpha$ emitters at
$z\sim 3.4$ and 4.5, Hu et al.\ (1998) estimated a surface density of
Ly$\alpha$ emitters of $\sim 13,000/{\rm unit}\ z/\sq^{\circ}$ down to an
emission flux of $\sim 10^{-17}$ ergs cm$^{-2}$ sec$^{-1}$.  These estimates
were found to be consistent with a complementary deep spectroscopic study,
where the slit provided limited spatial sampling, but wide wavelength
coverage at somewhat greater sensitivity, and recovered 4 Ly$\alpha$ emitters
with redshifts from $3.05\to 5.64$.  Preliminary results from the $z\sim5.7$
searches suggest number densities perhaps a factor of 6 lower than the value
for the lower redshift systems given by Hu et al.\ (1998).

\begin{figure}
\plotone{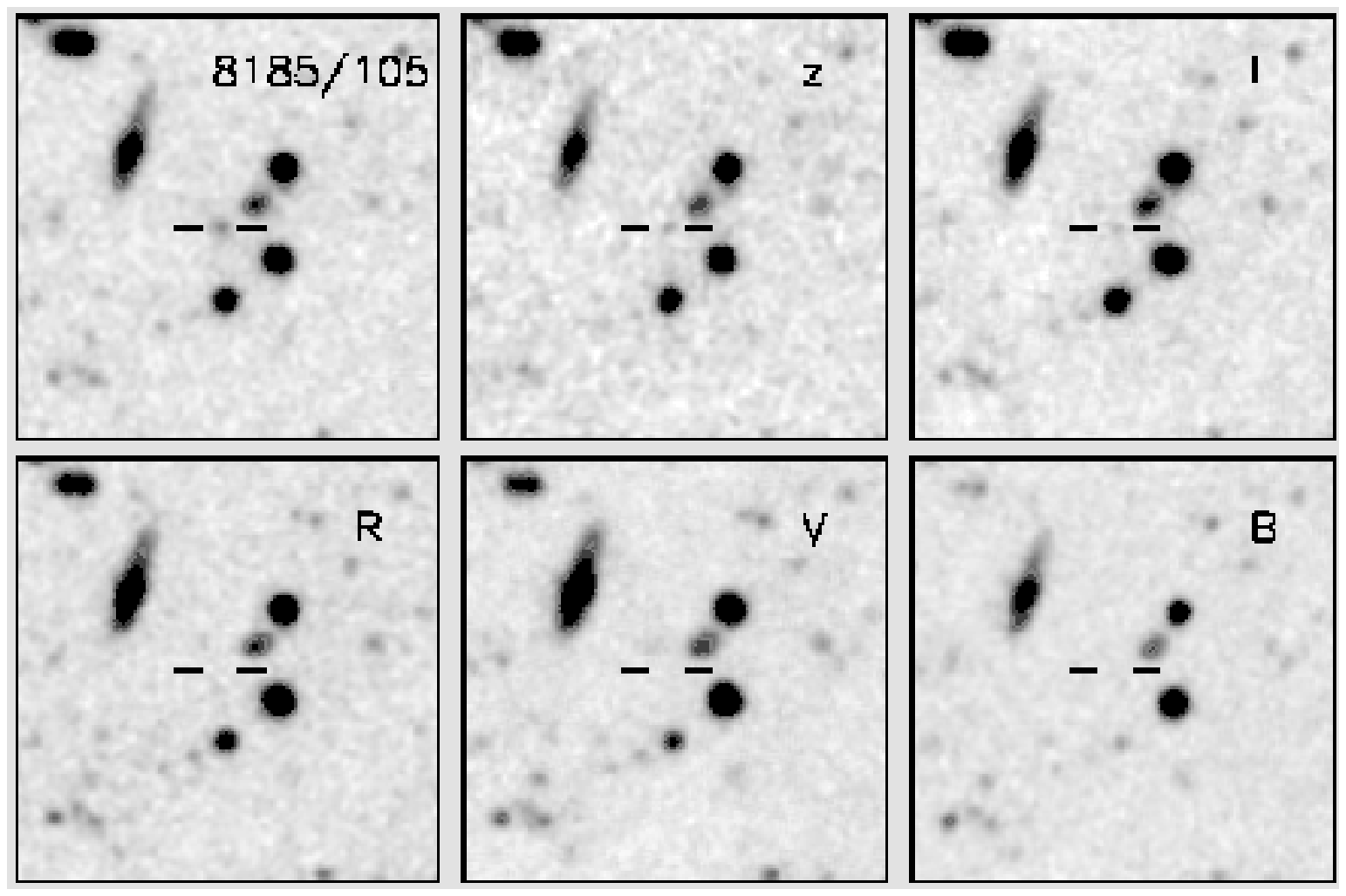}
\caption{\small Multicolor $B$, $V$, $R$, $I$, $Z$, and narrowband 8185/105
\AA\ images taken with LRIS on Keck of the $z=5.74$ object, SSA22-HCM1.  Each
panel is 30$''$ on a side.  The strong contrast of this object in the 8185
\AA\ narrowband filter can be seen, in comparison with the $Z$ band, which
samples the continuum in a region free of emission.  SSA22-HCM1 is an $R$
`dropout' and the object is absent at $B$, $V$, and $R$ in these extremely
deep Keck LRIS images.}\label{fig:1}
\end{figure}

\section{Spectroscopic Searches}

The surface number density estimates for Ly$\alpha$ emitters imply that
sufficiently deep spectroscopic exposures may intercept some of these
systems, at least over the lower redshift ranges.  In fact, the first
published $z>5$ galaxy, RD-1, was a serendipitously discovered Ly$\alpha$
emitter at $z=5.34$ reported by Dey et al.\ (1998) during deep long-slit
observations of a neighboring object in the field, with both emission line
and break identified in the spectra, and with the break confirmed by
broad-band filter imaging.  For RD-1, estimates of the continuum magnitude
above the Lyman break are $\sim 26.3$ AB mags 
using the broad-band imaging data, with emission-line
flux = $3.5\times 10^{-17}$ ergs cm$^{-2}$ s$^{-1}$, $W_{\lambda} = 600$
\AA\ (Dey et al.\ 1998).  The success of the spectroscopic searches depends
on the surface number density at the redshifts and fluxes being probed.
Equivalent width estimates, and particularly continuum magnitudes for faint
objects measured in dispersed modes have larger associated errors than
estimates tied to standard broadband filter photometry. Higher sensitivities
trade off against small area coverage, and for slit spectroscopy there is the
additional issue of whether with chance superpositions sample objects fall
only partly within the slit.  The additional problem in using this method as
a search technique is that it is difficult to assess the fraction of emitters
missed because they lie in regions of strong nightsky lines, and that for
such cases confirming observations can be extremely hard to make.

An interesting variant of this search technique is to use slitless
spectroscopy (Chen et al.\ 1999).  Incomplete object sampling by a slit is
removed, but the general problem of overlapping galaxy spectra must be dealt
with.  Using extremely deep STIS parallel exposures from HST, Chen et
al.\ (1999) identify a Ly$\alpha$ emitter at a probable redshift of 6.68.
At $z=6.68$ Ly$\alpha$ falls in regions of strong OH lines, which makes
confirmation with ground-based observations extremely challenging.

\begin{figure}
\plotone{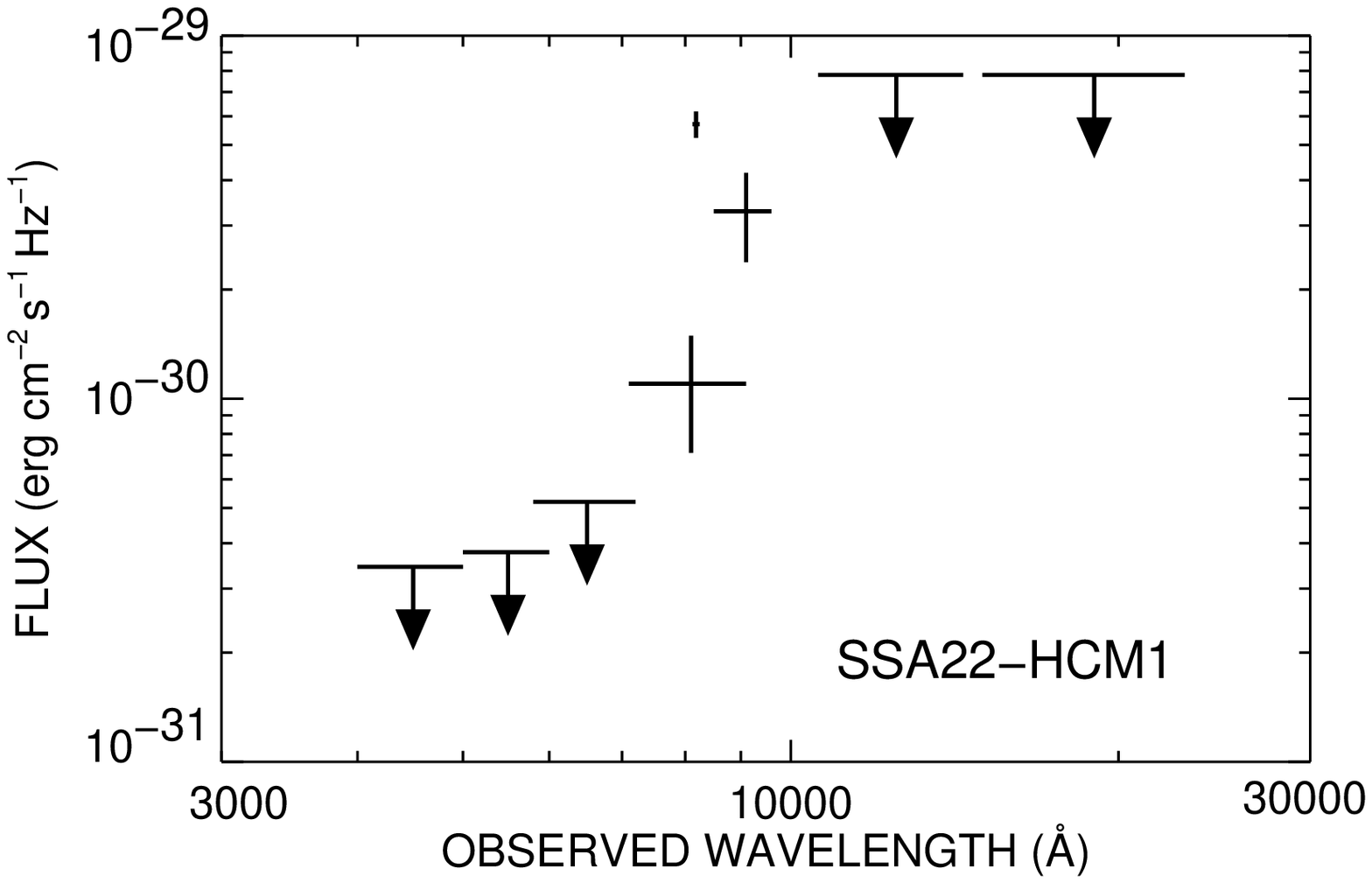}
\vspace{-0.15in}
\caption{\small Spectral energy distribution for SSA22-HCM1 obtained using
LRIS, and showing the bandwidths and errors for measurements in $B$, $V$, $R$,
$I$, narrowband 8185/105 \AA\ (emission), RG850 (line-free continuum longwards
of the emission), $J$, and $H+K'$. Significantly, the object has no flux in
the $R$ band, where strong Lyman $\alpha$ forest absorption is expected to be
present. The strong break ($2\sigma$ upper limit of 0.23 across the line) 
combined with the strong emission line is a signature for Ly$\alpha$.}
\label{fig:2}
\end{figure}

\begin{figure}
\plotfiddle{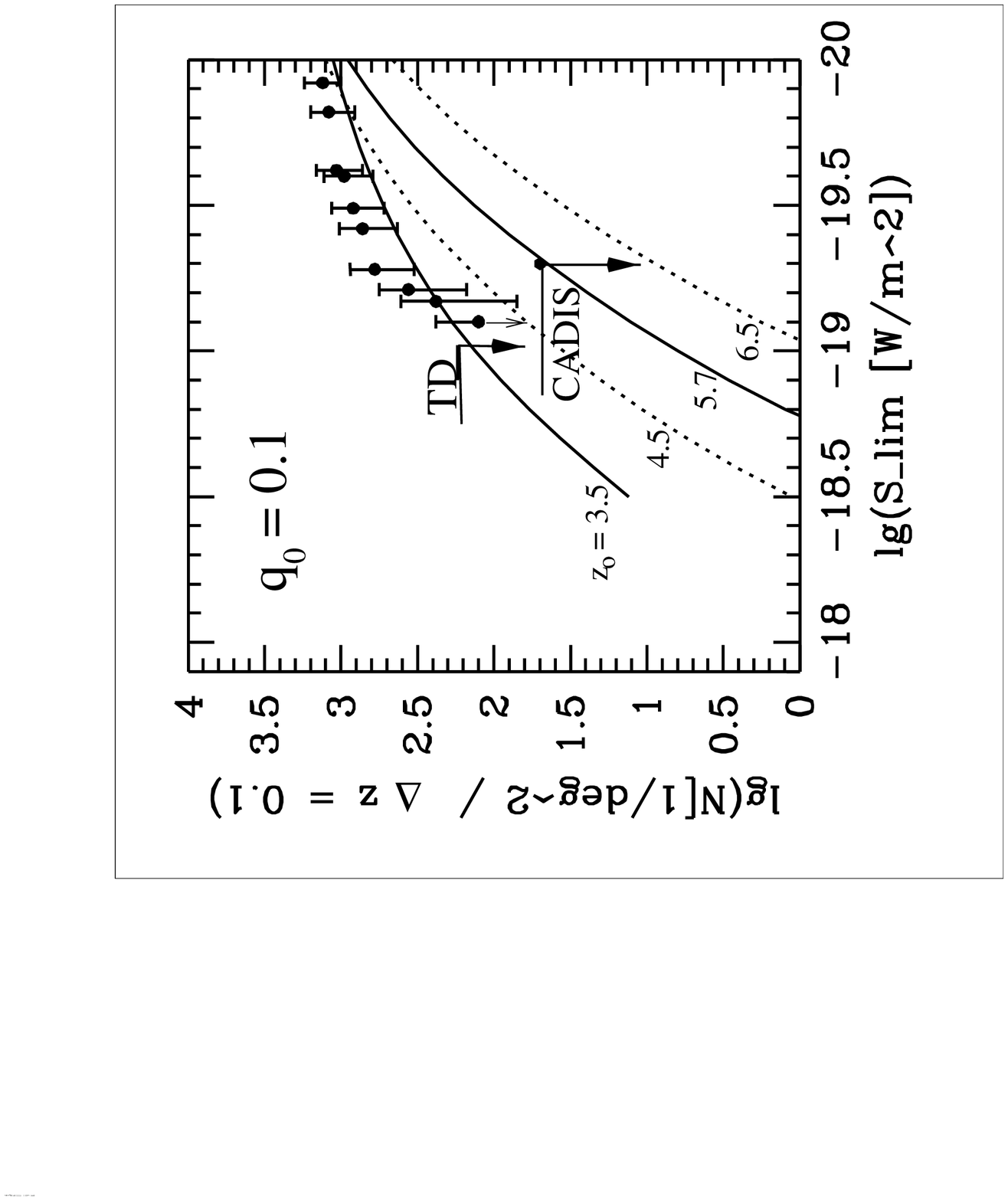}{250pt}{-90}{58}{58}{-235}{295}
\caption{Theoretical models (Thommes \& Meisenheimer 1999, in preparation)
for the surface density of high-$z$ Ly$\alpha$ emitters as a function of flux
and redshift.  The plotted points show the fits to the $z=3.43$ emitters
(Cowie \& Hu 1998).  Arrows show limits from non-detections by earlier
long-slit surveys of Thompson \& Djorgovski (1995) and of the current limits
for the Fabry-Perot surveys by the CADIS group (Thommes et al.\ 1998).  The
main qualitative points to note are: (1) the surface density of emitters
falls off dramatically as a function of limiting survey flux and (2) reaching
the redshift $z>6$ population critically requires getting down to below
$10^{-17}$ erg cm$^{-2}$ s$^{-1}$.  At current sensitivities coverage of
fairly wide areas is desired.
\label{fig:3}}
\end{figure}

\section{Color Breaks}

The strong depression of the continuum below the redshifted Lyman break caused
by the numerous neutral hydrogen absorbers dubbed ``the Ly$\alpha$ forest''
is the most notable feature in the galaxy continuum light of very
high redshift systems, where the extreme faintness of these objects,
typically $>26$ mags (AB) above the break, precludes the use of 
detailed absorption features to estimate the redshift.  This method has been
most successfully applied (e.g., Fern\'andez-Soto et al.\ 1999) in the 
Hubble Deep Field (HDF; Williams et al.\ 1996),
where the high precision of the photometric measurements permits robust
detection of a break signature, and where the availability of photometric
data at longer wavelengths (NICMOS infrared observations) allows
discrimination of Lyman break galaxies from red objects.  HDF 4-473.0 
(Weymann et al.\ 1998), shown to be a `$V$\/ dropout' by the absence of 
detectable flux in the WFPC2 F606W filter, combined with a roughly
flat $f_{\nu}$ spectrum out through 1.6$\mu$m, was confirmed as a
$z=5.60$ galaxy through deep LRIS spectra on Keck.  An emission line with
flux $\sim10^{-17}$ ergs cm$^{-2}$ s$^{-1}$ and $W_{\lambda}=300$ \AA,
identified as Ly$\alpha$, was used to establish the redshift. The magnitude
above the break is $\sim26.6$ AB mags.

A second color selected target, HDF 3-951.1 and 3-951.2, was identified
by the color break, and placed at an estimated redshift $z=5.34$
(Spinrad et al.\ 1998) on the basis of the break location.  For this
close pair, no emission features are detected (Fig.\ 2 of Spinrad
et al.\ 1998), with upper limits on possible Ly$\alpha$ more than
a factor of 10 below the observed flux of HDF 4-473.0  Spinrad et al.\ (1998)
consider the possibility that the absence of Ly$\alpha$ emission might be
associated with its brighter continuum.  Because studies at this depth and
precision of color measurement are only available for a small region of sky
($\sim5$ arcsec$^2$ each for the HDF and HDF South), key questions one would
like to address from the HDF color-break selected galaxies are:
(1) How typical are these properties of the high-redshift galaxies; what
are the consequences for future detections? 
(2) Are the bright high-redshift galaxies devoid of Ly$\alpha$ emission? 

\section{Properties of the $z>5$ Galaxies}

Figs.~\ref{fig:1} and \ref{fig:2} show images and a spectral energy
distribution (SED) for the $z=5.74$ galaxy, SSA22-HCM1 (Hu et al.\ 1999).
This object is an `$R$' dropout, and is notably absent in deep Keck LRIS
images in $B$, $V$, and $R$, which reach $1\sigma$ limits of $B$=28.3,
$V$=28.2, and $R=27.8$ for a $2''$ diameter aperture.  The SED shows both the
strong Lyman break and the Ly$\alpha$ emission feature (flux=$1.75 \times
10^{-17}$ ergs cm$^{-2}$ s$^{-1}$; $W_{\lambda}$ = 175 \AA), which was
confirmed with deep LRIS spectroscopy.  The increased contrast in the
appearance of the object as detected in the 8185 \AA\ narrowband filter
compared with the line-free $Z$ band filter around 9200 \AA\ may be noted.
For these measurements we use fluxes measured in the narrowband filter,
instead of values recovered from spectroscopic extractions, for greater
precision.  Aperture 1$\sigma$ errors are estimated from laying down random
apertures away from identified sources, since in the deep exposures magnitude
limits are set by the background faint source population (Fig.~\ref{fig:1}).
The measured continuum above the break, 25.5 mags (AB), is a magnitude higher
than the Weymann et al.\ (1998) observed estimate for HDF 4-473.0, and lies
between our estimated continuum magnitudes of 24.9 and 25.7 (based on our
deep $Z$ band observations of the HDF) for the 3-951 pair.  SSA22-HCM1 is the
brightest of the $z=5.7$ Ly$\alpha$ emitters surveyed to date, and in
contrast to 3-951 has strong emission.  However, it appears that more typical
emitters at these redshifts have properties like 4-473.0 (Hu et al.\ 1999, in
preparation).  We can summarize the expected properties for
Ly$\alpha$-emitting galaxies above $z\sim 6$, based on the current $z\sim
5.7$ survey, as:  Ly$\alpha$ fluxes $\sim10^{-17}$ ergs cm$^{-2}$ s$^{-1}$,
equivalent widths of a few 100 \AA, and continuum magnitudes fainter than
$\sim26.5$.

Efforts to model the high-redshift emitters have begun (e.g., Thommes 1998,
Haiman \& Spaans 1999), and we can use these as a starting point to estimate
the detection limits that will be required for working at higher redshifts.
Fig.~\ref{fig:3} from Thommes (1998) shows the expected surface number
density of Ly$\alpha$ emitters for four redshift intervals ($z=3.5, 4.5, 5.7,
{\rm and} 6.5$) corresponding to spectral regions free of strong nightsky
lines.  Overplotted are the data from Cowie \& Hu (1998) for $z=3.4$, and
other upper limits.  It can be seen that, in agreement with the properties
summarized above, surveys will need to exceed $5\sigma$ detection limits of
$10^{-17}$ ergs cm$^{-2}$ s$^{-1}$ ($10^{-20}$ W m$^{-2}$), and preferably
cover wide areas.

\section{($R-Z$) Color Selection}

\begin{figure}
\plotone{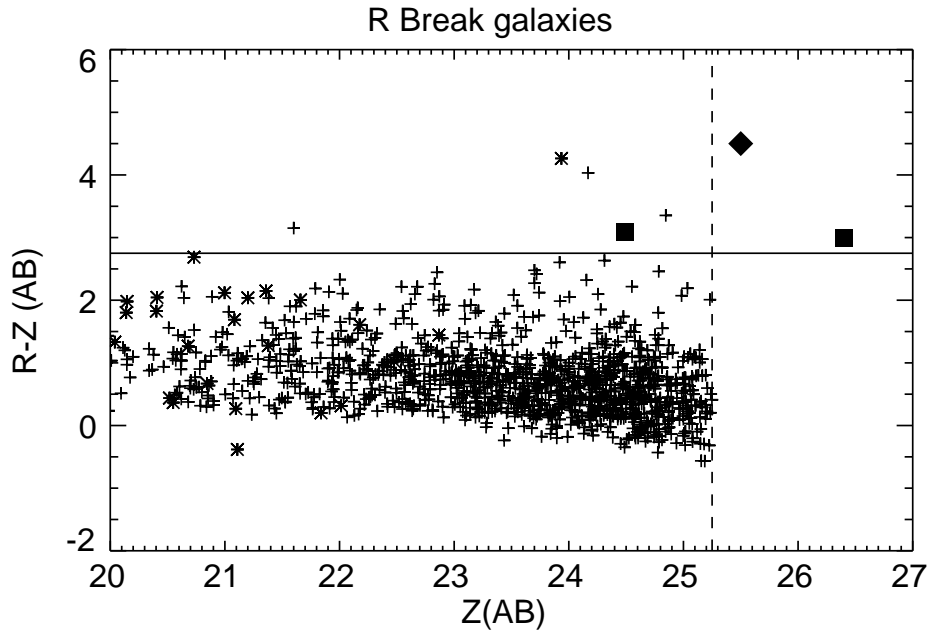}
\vspace{-0.45in}
\caption{\small ($R-Z$) colors of galaxies in the HDF and SSA22 fields. 
The dashed vertical line shows the $5 \sigma$ magnitude limit in $Z$. The 
filled squares show the colors of (both components of) HDF 3-395 and HDF
4-473.0, and the filled diamond shows the measured ($R-Z$) color for
SSA22-HCM1.  The solid line at ($R-Z$) = 2.75 provides an arbitrary
divider for very red objects.  Known stars (indicated with asterisks) can
lie in the region above the line.  Galaxies populating the upper part of
the plot include both dust-reddened objects and high-$z$ continuum break
objects.}
\label{fig:4}
\end{figure}

Because SSA22-HCM1 is so bright in $Z$ it is interesting to examine color
statistics for objects in ($R-Z$) vs. $Z$.  Fig.~\ref{fig:4} shows the
color distribution of objects in the HDF and SSA22 fields over a sample
area of 75 arcmin$^2$. Magnitudes are measured on the AB
system using $2''$ diameter corrected apertures. 
The fiducial color boundary used
here of $(R-Z)> 2.75$ lies marginally above the range of possible M star 
contaminants, which might be distinguished by compactness criteria.
At faint magnitudes the depth of the $R$ exposure dominates the
errors in the $(R-Z)$ color measurement, which can be quite large.  
Filled symbols indicate known
$z>5$ galaxies (or in the case of HDF 3-395.1 and 3-395.2, the summed
magnitudes of the object pair). In the case of HDF 4-473.0, the
colors are estimated from deep WFPC2 and NICMOS photometry
(Weymann et al.\ 1998).
The $5\sigma$ $Z$ band criterion is not sufficiently deep to include
SSA22-HCM1, the $z=5.60$ galaxy HDF 4-473.0, or the $z=5.34$ emission-line
galaxy RD1 (based on the estimated $I(AB)=26.1$), but the threshold is only
a few tenths of a magnitude from reaching the brighter two objects.
Thus, it appears that 
the use of deep $Z$-band filter imaging and $R-Z$ color selection on the
large telescopes may be a possible way to extend methods for
selecting candidates to higher ($z > 5.7$) redshifts for ground-based
studies, although the background sky brightness near 9200 \AA\ is high.
This could be used in combination with a near-IR color measurement to
distinguish break objects from highly reddened galaxies.

\section{Evolution of the Continuum Break}

\begin{figure}
\plotone{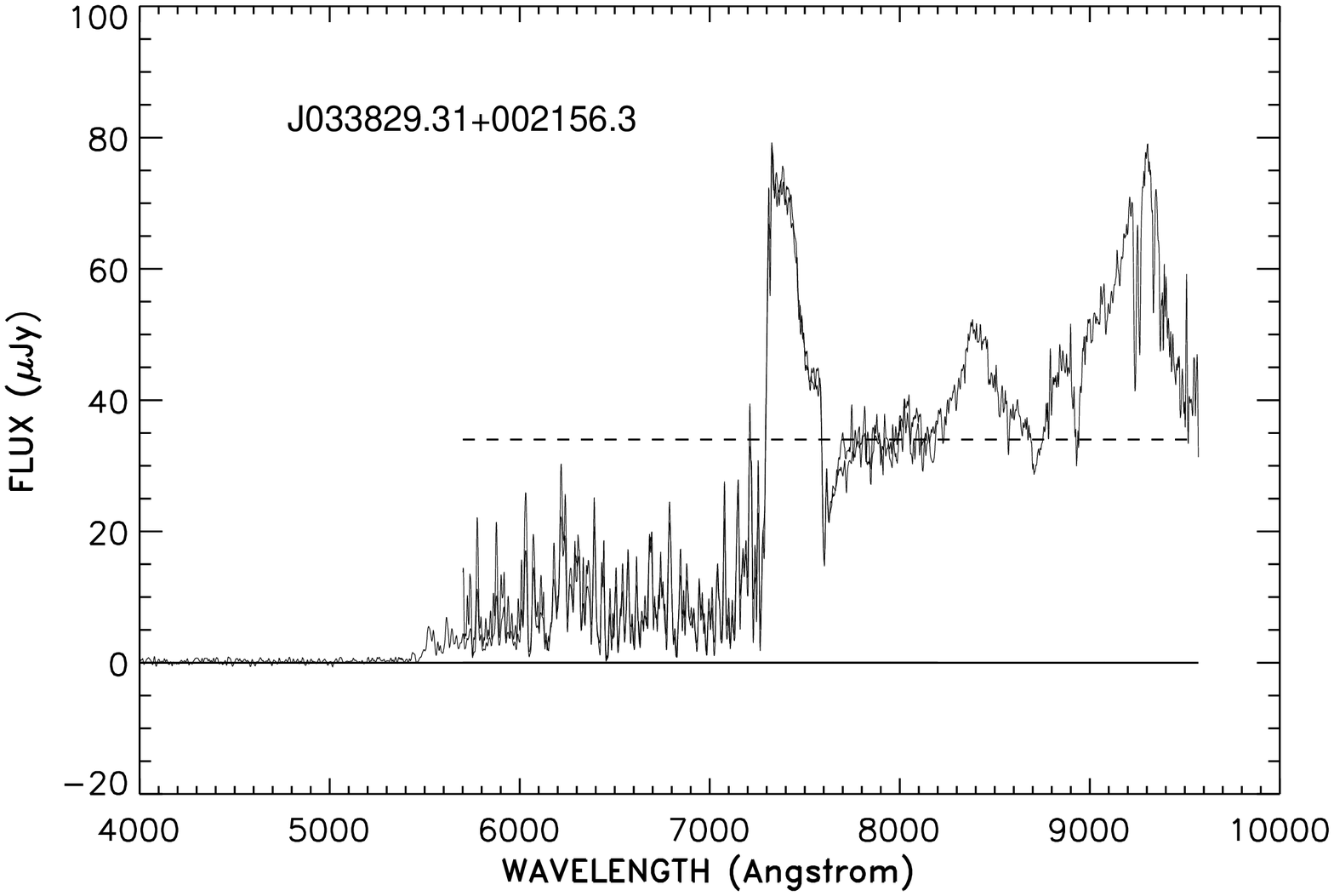}
\vspace{-0.25in}
\caption{\small The Lyman break strength at high redshift is estimated
to be a factor of 4 from LRIS spectra (Songaila et al.\ 1999)
of the $z=5.00$ Sloan Survey quasar (Fan et al. 1999).}
\label{fig:5}
\end{figure}

\begin{figure}
\plotone{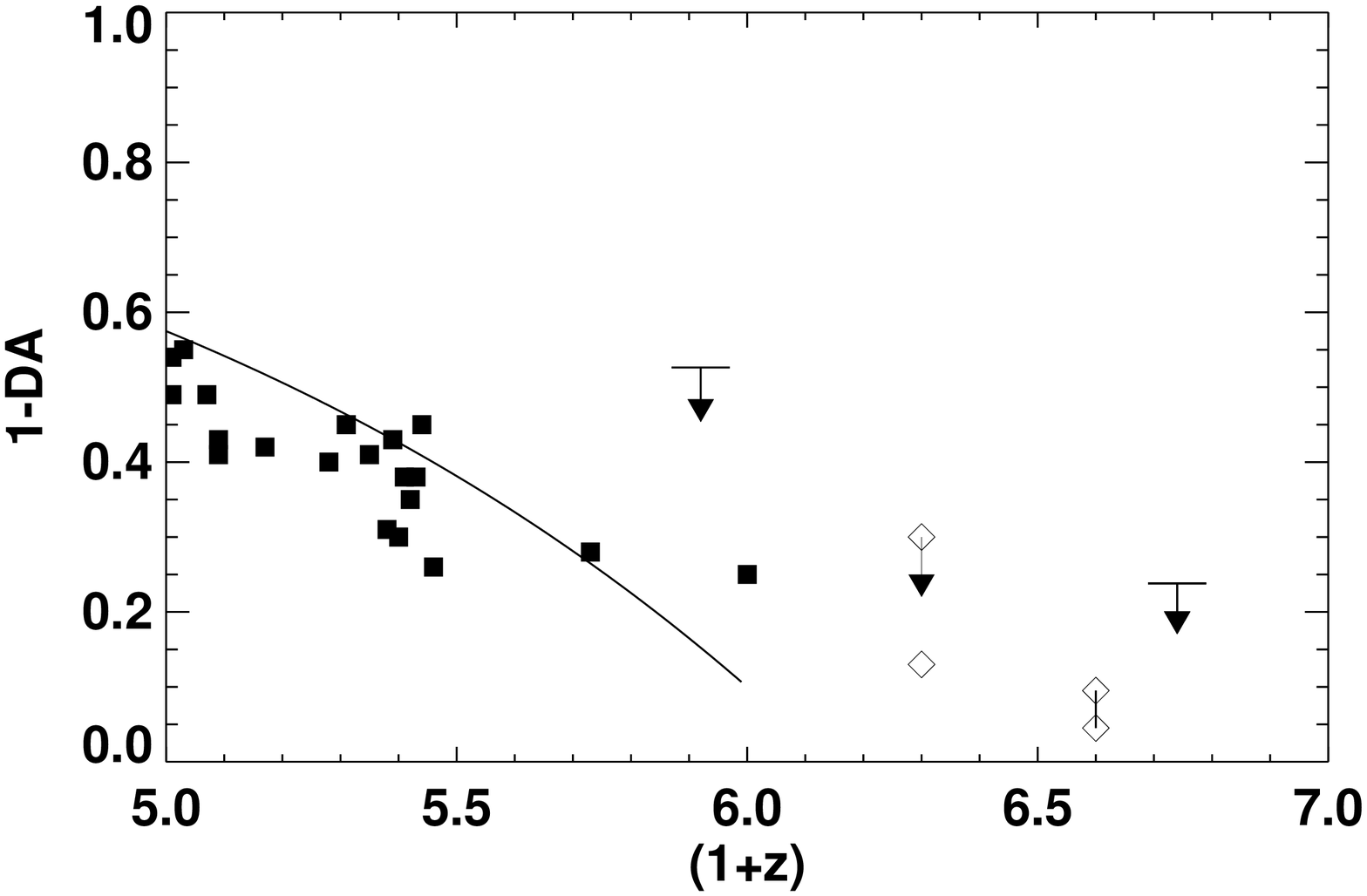}
\vspace{-0.25in}
\caption{\small The measured Lyman break strength (1-$D_A$)
at high redshifts for quasars ([{\it filled squares}] Schneider et al.\ 1991; 
Kennefick et al.\ 1995; Songaila et al.\
1999) and high-redshift galaxies HDF 4-473.0 and 3-395({\it open diamonds}).  
Upper limits are indicated for the $z=4.92$ lensed galaxy (Franx et al.\ 1997)
and for SSA22-HCM1 at $z=5.74$ (Hu et al.\ 1999). The solid curve shows
Zhang et al.'s (1997) extrapolation using a Haardt-Madau (1996) spectrum.
}
\label{fig:6}
\end{figure}

The new $z\sim5$ quasars discovered by the Sloan Survey (Fan et al.\ 1999)
provide a means of evaluating the expected continuum depression below
the Ly$\alpha$ line at high redshifts. Because of the rapid increase in the gas density with redshift, it is expected
that, at the higher redshifts, there will be a sizable break. In 
Fig.~\ref{fig:5} we show a spectrum of the newly discovered $z=5$ quasar,
J033829.31+002156.3, for which we determine a factor of 4 break across
the emission line. By comparison, Zhang et al.\ (1997) find, at $z = 5$, for 
$H_{\circ} =
50~{\rm km\ s^{-1}\ Mpc^{-1}}$, $\Omega_b = 0.06$\ and a Haardt-Madau
(1996) spectrum and ionizing flux, that the spectrum of a $z = 5$\ quasar will
have an average flux in the region between $1050$ \AA\ and $1170$ \AA\ that is
10\% of the continuum value which would be present in the absence of Lyman
alpha scattering.
A plot of the continuum break across Ly$\alpha$ for various $z>4$
quasars (Schneider et al.\ 1991; Kennefick et al.\ 1995) and galaxies is shown
in Fig.~\ref{fig:6}.  $D_A$ is the index (Oke \& Korycanski 1982)
$$D_A = \left\langle 1 - {{f_{\nu}({\rm observed})} \over {f_{\nu}({\rm
continuum})}} \right\rangle$$
in the rest-frame wavelength range $1050$ \AA\ to $1170$ \AA.

\section{Future Investigations}

It is useful to consider possible directions for future studies.
A number of $z>4$ galaxies have been found using cluster lens
systems (e.g., Trager et al.\ 1997, Franx et al.\ 1997, Frye \&
Broadhurst 1998, Pell\'o et al.\ 1999), where amplification of
the source has increased its detectability.  In general, such 
discoveries do not provide information on the statistics, continuum
magnitudes, and base emission-line fluxes of the
high-redshift galaxies found, but for very well-studied cluster lens
systems (e.g., B\'ezecourt et al.\ 1999) it may be possible to recover such information. 
Typical magnification is about 1 magnitude (Smail et al.\/ 1999), and this 
may be a fruitful way to extend high-redshift searches.

The availability of new IR spectrographs for the 8--10-m. class
telescopes will make it possible to study [O\thinspace{\sc{ii}}]
associated with the $z<5$ systems (e.g. Egami 1997).  Such observations
may provide more information on the degree of dust extinction in
the high-redshift galaxies detected by their Ly$\alpha$.  However,
for $z>5$ systems [O\thinspace{\sc{ii}}] lies in the thermal IR
and cannot be readily studied from the ground.

For redshifts $z>5$, high equivalent width H$\alpha$ emission from
high ionization foreground extragalactic H{\sc{ii}} regions is a possible
contaminant which can mimic Ly$\alpha$.  Such systems may have no detectable 
continuum, and [N\thinspace{\sc{ii}}] suppressed to less than 2\% of H$\alpha$
(Stockton \& Ridgway 1998).  The incidence of these systems is not known,
but a few cases have been encountered in our $z\sim5.7$ and $z\sim6.5$
surveys --- and identified by wide wavelength spectra covering the
corresponding [O\thinspace{\sc{ii}}] and [O\thinspace{\sc{iii}}].
Extensive wavelength coverage is required to verify sources as Ly$\alpha$
emitters.

Searches for yet higher $z$ galaxies are of necessity forced into the near 
IR since
the light below $1216(1+z)~{\rm\AA}$\ will be essentially extinguished by the
strong Ly~$\alpha$\ forest blanketing or by H~I Gunn-Peterson (Miralda-Escud\'e
\& Rees 1998) at the redshifts
where the IGM was neutral.  We can look forward to detecting these objects
with the Next Generation Space Telescope.

\acknowledgments
This work was supported in part by NASA grants GO-6626.01-95A, GO-7266.01-96A, 
and AR-6377.06-94A from Space Telescope Science Institute, which is 
operated by AURA, Inc. under NASA contract. Results reported here include 
observations made as a Visiting Astronomer at the W. M. Keck Observatory,
which is jointly operated by the California Institute of Technology, the
University of California, and the National Aeronautics and Space Administration.

\end{document}